# Covalent pathways in engineering *h*-BN supported graphene

Bin Ouyang[1], J. Song[1, a)]

1. Department of Mining and Materials Engineering, McGill University, Montreal, QC, Canada

## Abstract

Cross-planar di-vacancies (CPDVs) within stacked graphene hexagonal boron nitride (*h*-BN) heterostructures provide stabilized covalent links to bridge adjacent graphene and *h*-BN sheets. It was shown that the CPDVs serve as focal points for cross-planar atom transport between graphene and *h*-BN, and the chemical nature of interlayer links along with associated cross-planar migration pathways at these defects can be predictively manipulated through modulation of the chemical environment and charge engineering, to achieve consistent B or N doping and simultaneous healing of graphene. The present study proposed a viable approach integrating irradiation, chemical and charge engineering, to produce high-quality graphene with tunable electronic and electrochemical properties, using the *h*-BN substrate.

**Keywords:** graphene heterostructure, *h*-BN substrate, impurity doping, self-healing, cross-planar defects, chemical potential, charge engineering

[a)] Author to whom correspondence should be addressed. E-Mail: jun.song2@mcgill.ca

**Introduction:**

Interface coupling between graphene and its insulating isomorph, hexagonal boron nitride (*h*-BN) provides interesting possibilities for the synthesis and engineering of graphene-based two-dimensional nanomaterials. The predominant Van der Waals interactions between them enables an atomic sharp interface that minimizes dangling bonds and charge traps, making *h*-BN as a promising substrate for high-quality graphene devices. Meanwhile the accompanying electronic coupling between graphene and *h*-BN leads to compelling physical phenomena, such as breakage of time reversible symmetry[1-4], commensurate to incommensurate transition[5-7], and Hofstadter butterfly[4, 8, 9], promising numerous ways to manipulate graphene devices through periodic potential.

Besides the above coupling effects derived from the long-range dispersive interlayer interactions, the interplay between graphene and *h*-BN may also be affected by discrete covalent connections. Telling et al.[10] demonstrated the existence and ground states of cross-planar di-vacancies (CPDVs), namely Wigner defects in graphite. These defects, introduced via high-energy (e.g., irradiation) and high-temperature processes[11-13], induce local three-dimensional (3D) reconstruction and bridges adjacent atomic sheets in graphite through covalent bonds. Given the close structural resemblance between graphene and *h*-BN, and sp2 hybridization bonding in both materials, CPDVs are also expected to exist at the graphene/*h*-BN interface. These defects have great implications for the *h*-BN supported graphene. Aside from the well expected strengthening attributed to the interlayer covalent bonding as hinted in previous studies on graphite and carbon nanotubes[12, 14-16], the CPDVs create stabilized links between graphene and the two non-equivalent sublattices in *h*-BN. These links (i.e., C-B or C-N) are of distinct energy states, bond polarization and electronic structures, and can act as potential focal points for structural evolution and electron doping of graphene. In this paper, we present the first systematic study of CPDVs at the graphene/*h*-BN interface, and show that those CPDVs can serve as effective channels for cross-planar atom transport. Moreover, through modulation of the chemical environment and charge state, the cross-planar transport can be manipulated to yield controlled B or N doping of graphene. Simultaneously with the doping of graphene, the CPDV converts to an in-plane di-vacancy (DV) in *h*-BN and the cross-planar covalent bond gets annihilated, resulting in *healing* of vacancies in graphene. These findings promise viable routes to manipulate defect evolution to enable composition and electrochemical engineering of *h*-BN supported graphene.

**Results and Discussions:**

A CPDV at the graphene/*h*-BN interface is formed when two single vacancies (SVs) at adjacent sheets coalesce. For the coalescence to occur, the two SVs need to be in close vicinity of each other to enable the overlap of cross-planar dangling orbitals so that bond reconstruction can be induced by lattice fluctuation. Figure 1 shows the eight possible CPDV configurations (immediately before the coalescence of SVs) at the graphene/*h*-BN interface, identified via DFT calculations[11]. For simplicity, we adopt a notation similar to what Telling et al. used in Ref. [10] to distinguish different CPDV complexes. As illustrated in Fig. 1, a CPDV configuration is denoted as $V_2^i(C_\alpha X)$ where the superscript $i$ = 1 or 2 indicate that the constituent vacant sites are 1st or 2nd nearest interplanar neighbors with each other, $C_a$ refers to the missing C atom of which $a$ represents two sublattices in graphene layer that could be either $\psi$ (sublattice sits on top of B atom) and $\varphi$ (sublattice sits on top of the boron nitride hexagon center), and $X$ stands for the missing atom in *h*-BN (either B or N atom). The constituent single vacancies are denoted as $SV_C$, $SV_B$ and $SV_N$, referring to the single C vacancy in graphene, and single B and N vacancies in *h*-BN respectively.

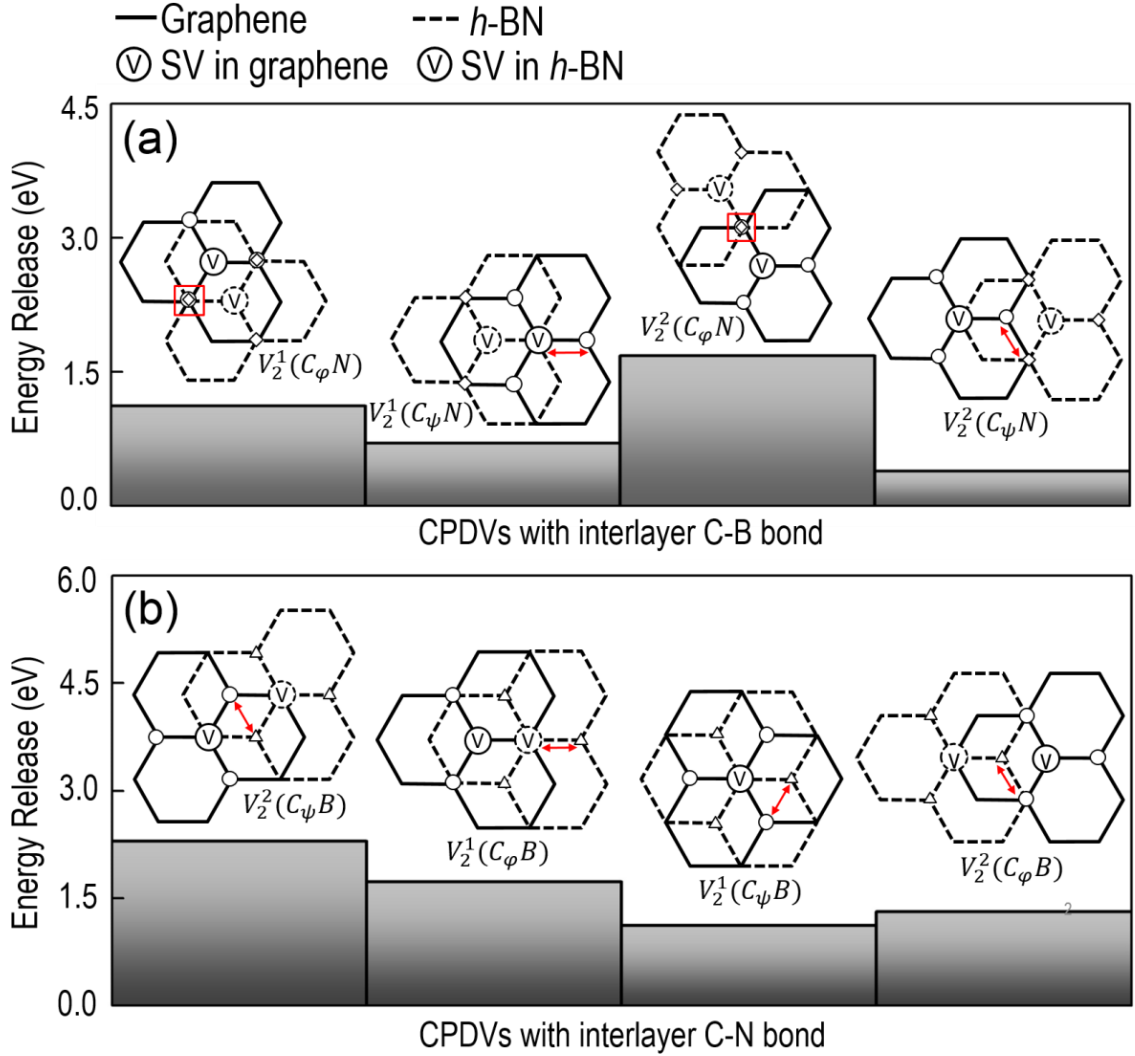

**Fig. 1** (Color online). The corresponding energy release associated with the formation of CPDVs with (*a*) interlayer C-B links and (*b*) interlayer C-N links. The insert figures illustrate the local configurations of corresponding SV couples prior to the CPDV formation. The atom pairs responsible for the interlayer links are indicated by red squares or left-right arrows. Dangling C, B and N atoms neighboring CPDVs are indicated by open circles, open diamonds and open triangles respectively.

The ground states of those CPDVs are obtained through structural optimization in DFT calculations. The coalescence from two interlayer SVs into a CPDV leads to sizable energy release, clearly noted in Fig. 1. Figs. 2a and 2b show the relaxed atomic configurations of two representative CPDVs with C-B covalent bonds, i.e., $V_2^1(C_\psi N)$ and $V_2^2(C_\psi N)$ where graphene and *h*-BN sheets are bridged by interlayer C-B bonds. We can note that in both cases the interlayer bond induces considerable basal shearing and local buckling at the CPDV. Particularly for $V_2^1(C_\psi N)$, the interlayer bond produces a basal shift of 0.17Å along the armchair (AC) direction, and displaces the C and B atoms away from their corresponding host atomic sheets by 0.65Å and 1.24Å respectively (cf. Fig. 2a), while for $V_2^2(C_\psi N)$, the interlayer bond results in a basal shift of 0.38Å along the zigzag (ZZ) direction, and displaces the C and B atoms away from their corresponding host atomic sheets by 0.79Å and 0.94 Å (cf. Fig. 2b). Similar phenomena are also observed for CPDVs with interlayer C-N bonds, also illustrated by the two representative cases, i.e., $V_2^1(C_\psi B)$ with displacements of C and N atoms being respectively 1.00Å and 1.01Å and a basal shift of 0.19Å along AC direction, and $V_2^2(C_\psi B)$ with displacements of C and N atoms being respectively 0.89Å and 1.11Å and a basal shift of 0.51 Å, shown in Figs 2c and 2d respectively. These distortions are identified to be of direct relevance in determining the energetics of CPDVs (see Supplementary Information).

With the large local geometry modification, the CPDV necessarily induces significant perturbation in the spatial partial charge distribution. From the electronic structures deduced from DFT calculations, the Scanning tunneling microscope (STM)[17, 18] images of the aforementioned CPDVs are simulated, shown alongside with the corresponding atomic configurations in Fig. 2. We can see that CPDVs induce significant contrast in those simulated STM images. Meanwhile we can also note that the contrast is strongly correlated with the local geometrical characteristics associated with the CPDV, suggesting that CPDVs can be readily identified and distinguished in experiments employing STM.

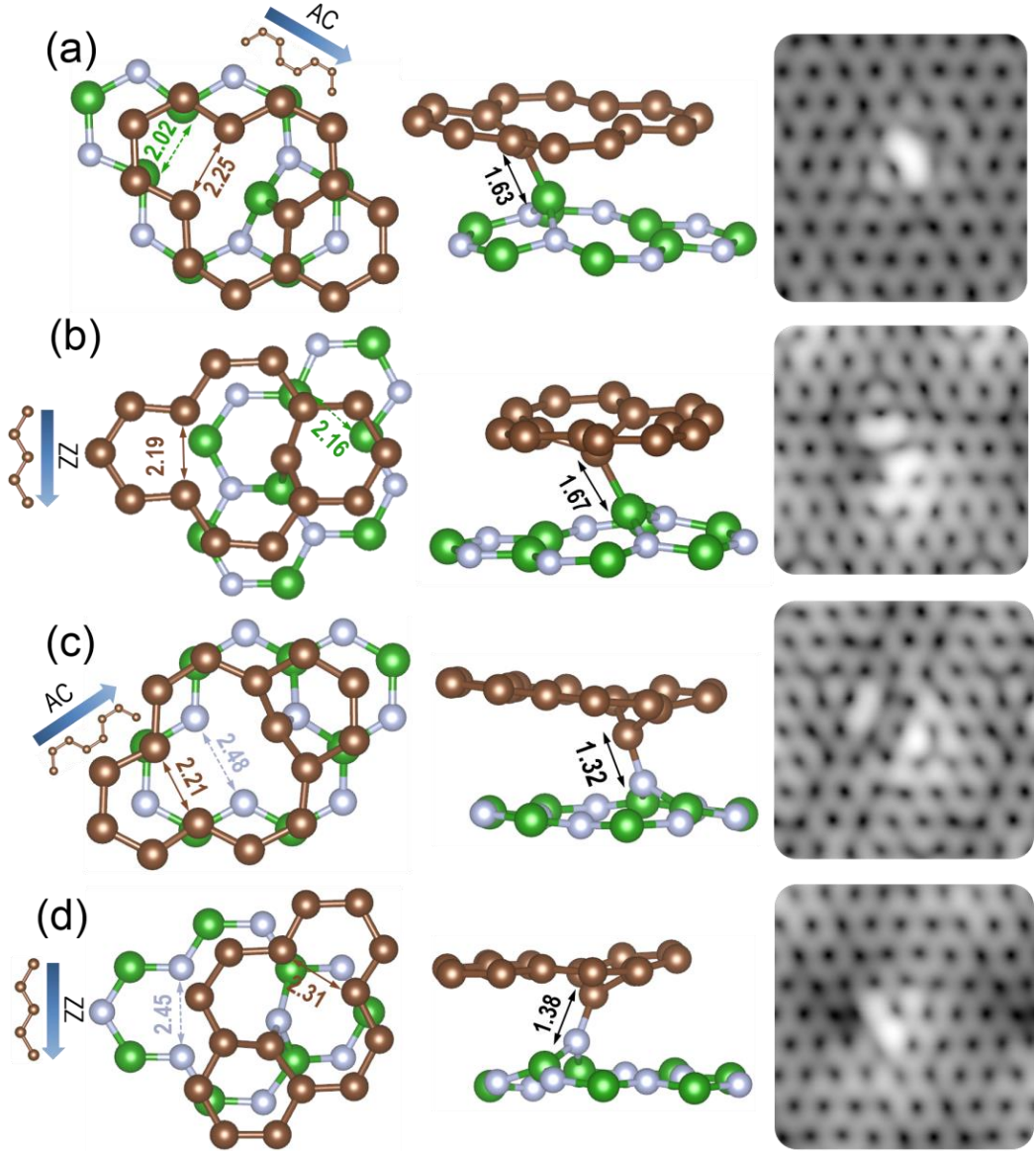

**Fig. 2** (Color online). The top and side views of the ground-state configurations of several representative CPDVs: (*a*) $V_2^1(C_\psi N)$, (*b*) $V_2^2(C_\psi N)$, (*c*) $V_2^1(C_\varphi B)$ and (*d*) $V_2^2(C_\varphi B)$. The subfigures in the rightmost column are the simulated STM images of those CPDVs. In the STM simulation, the partial charge range from $E_F - 1.5\text{eV}$ to $E_F$ (with $E_F$ being the corresponding Fermi energy) is considered to capture the defective charge state. C, B and N atoms are colored dark brown, green and silver respectively.

Locally at a CPDV, the adjacent graphene and *h*-BN sheets are distorted towards each other. In particular, the two atoms (i.e., C and B or C and N) that constitute the interlayer links are considerably displaced towards the opposite layers. These features would presumably aid cross-planar kinetics. In this regard, we systematically examined the cross-planar migration pathways at CPDVs combing DFT with transition state theory (TST)[19-22]. In our discussion below, $V_2^1(C_\psi N)$ and $V_2^1(C_\varphi B)$ are selected to represent CPDVs with interlayer C-B and C-N bonds respectively. The possible cross-planar migration scenarios and associated minimum energy paths (MEPs) at $V_2^1(C_\psi N)$ and $V_2^1(C_\varphi B)$ are shown in Fig. 3. There are two possible routes for cross-planar migration at a CPDV with the interlayer C-X (X = B or N) bond, i.e., A) atom transport towards graphene, resulting in X-doping of graphene and a planar divacancy in h-BN (DV$_C$), or B) atom transport towards h-BN, resulting in C-doping of h-BN and a planar divacancy in h-BN (DV$_{BN}$), both yielding further energy release compared to the CPDV. In particular for $V_2^1(C_\psi N)$, the migration route A exhibits an energy barrier of 0.76 eV and results in an energy release of 1.43 eV with respect to $V_2^1(C_\psi N)$, while the migration route B shows a much higher energy barrier of 2.79 eV and leads to a smaller energy release of 0.75 eV. Meanwhile for $V_2^1(C_\varphi B)$, the migration route B is both kinetically and energetically favored over route A, showing a lower energy barrier of 0.98 eV (than 1.57 eV in route A) and higher energy release of 4.84 eV (than 3.83 eV in route A). One thing to note is that cross-planar migration at a CPDV always results in energy release and is thus thermodynamically favored. The energy release is well expected by looking at the formation energy of the post-migration defect complex (denoted as $E_f^{PM}$ below), which can be roughly estimated from individual energetics data of defects in graphene and h-BN systems[23-30],

$$E_f^{PM} \approx \begin{cases} E_f[DV_{BN}] + E_f[C_X], & \text{Route A} \\ E_f[DV_C] + E_f[X_C], & \text{Route B} \end{cases} \quad (1)$$

where $E_f$[DV$_{BN}$] and $E_f$[DV$_C$] denote the formation energies of a divacancy in h-BN and graphene respectively, $E_f[C_X]$ denotes the formation energy of a substitutional impurity $X$ ($X=B$ or $N$) in graphene, and $E_f[X_C]$ denotes the formation energy of a C impurity that substitutes $X$ atom in $h$-BN.

Among the various cross-planar migration possibilities at a CPDV, of particular interest are the ones that lead to atom (B or N) transport towards graphene. As illustrated in Fig. 3, they result in B or N doping of graphene, which consequently can modify the electronic, chemical and magnetic properties as well as electrocatalytic activity of graphene[31-37]. In addition, those dopants fill in the otherwise vacant sites in graphene to help improve the lattice quality. This essentially leads to *healing* of graphene lattice, hinting a strategy to moderate defect density during doping of graphene. Nonetheless, these migration paths will be competing with the ones that result in atom transport towards $h$-BN which in sharp contrast further deteriorate the quality of the graphene. For instance, in the case of $V_2^1(C_\varphi B)$, the C doping of $h$-BN (i.e., route B) is both thermodynamically and kinetically favored over the N doping of graphene (i.e., route A). The above competition provides an interesting implication (and challenge) on the synthesis of impurity (i.e., B or N) doped graphene with enhanced lattice quality. To put things into context, below we discuss the formation of CPDVs and migration at CPDVs in possible experimental settings.

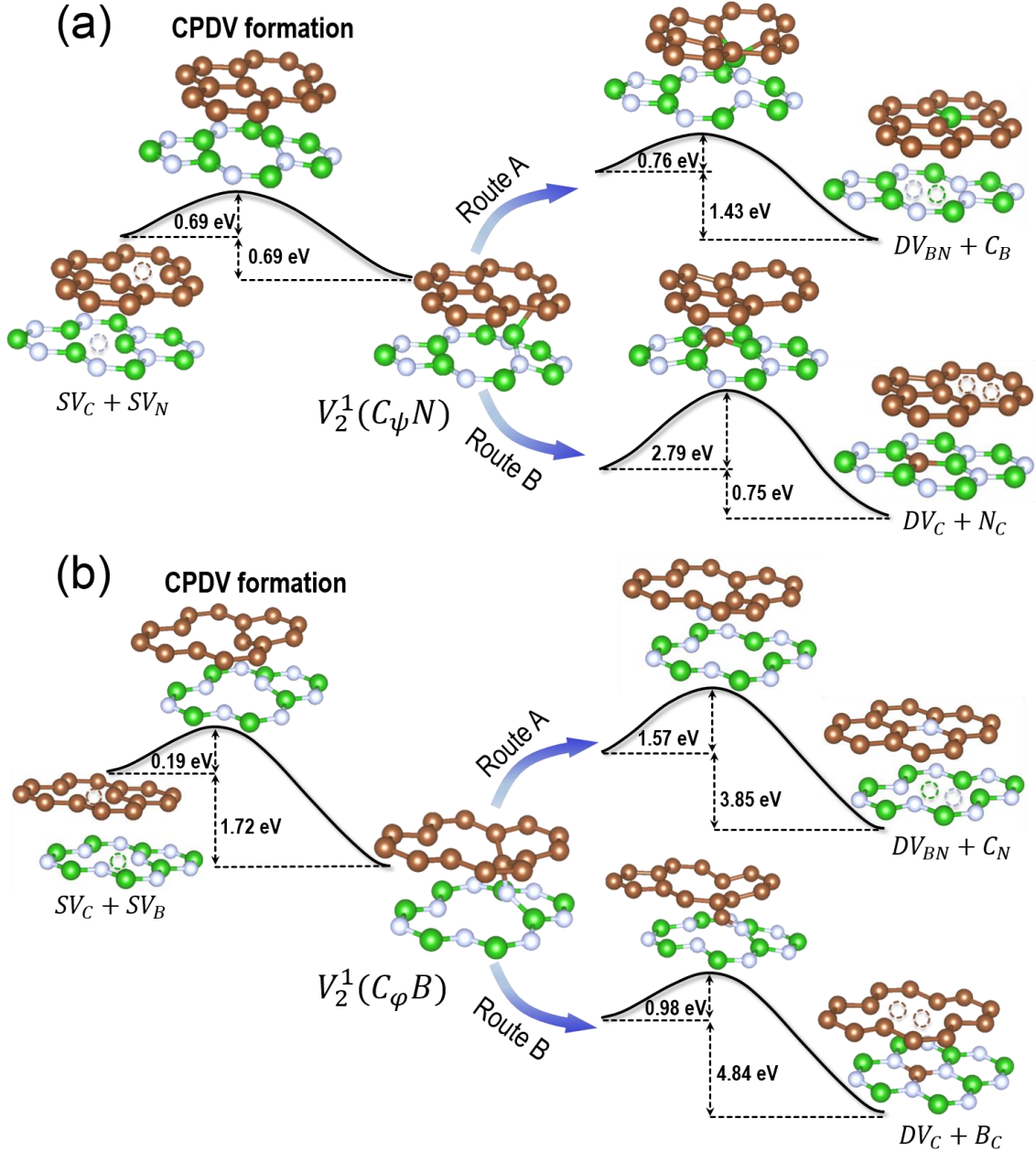

**Fig. 3** (Color online) The kinetics and reaction pathways of defect formation and subsequent cross-planar migration at two representative CPDVs: (*a*) $V_2^1(C_\psi N)$ and (*b*) $V_2^1(C_\varphi B)$, with the corresponding migration barriers indicated. C, B and N atoms are colored dark brown, green and silver respectively. The dashed circles indicate the SVs ($SV_C$: brown; $SV_B$: green and $SV_N$: silver) prior to the CPDV formation and divacancies (i.e., represented as SV couples) post migration.

One method often adopted in experiments to induce cross-planar defects in carbon-based nanomaterials is electron[38-40] or ion irradiation[41, 42]. The irradiation approach has been widely used to modify the structural properties of, e.g., carbon nanotubes[11, 12, 15, 16, 40-42], graphite[11, 14, 24] and other 2D structures[43-48]. During irradiation, the incoming particles deploy sufficient energy to cause large out-of-plane displacements of atoms and thus aid the formation of covalent interlayer links. With the graphene/*h*-BN system being structurally similar to carbon nanotubes and graphite, one can imagine that the irradiation method can also serve as an effective means to generate CPDVs between graphene and *h*-BN sheets. The different flavor is, however, that we would expect two different sets of interlayer links, C-B and C-N, rather than the C-C links in those pure carbon-based systems previously studied. Consequently the formation along with the bonding characteristics of the interlayer link of a CPDV in the graphene/*h*-BN system depend on the chemical environment (i.e., B-rich or N-rich) of the experiment. This dependence of chemical environment apparently also applies to those defect complexes that derive from CPDVs (e.g., the post-migration systems, cf. Eq. 1). Another important aspect of electron or ion irradiation is that it naturally brings charge into the material system treated[25, 48-50]. This modifies the charge state of the material system and introduces another dimension of influence to tune the energetics of resultant defect complexes. Accounting for the effects of chemical environment and charge state, we can formulate the formation energy, $E_f^q$, of a defect in the graphene/*h*-BN heterostructure as [25, 26, 51, 52]:

$$E_f^q = E_{tot}^q - N_B \mu_B - N_N \mu_N - N_C \mu_C - q E_F^q, \qquad (2)$$

where $q$ denotes the charge state of the system, $E_f^q$ and $E_F^q$ denote the total energy and Fermi energy of the system with a charge state $q$, $N_C$, $N_B$ and $N_N$ are the numbers of C, B and N atoms respectively, and $\mu_C$, $\mu_B$ and $\mu_N$ are the chemical potentials of C, B and N elements respectively..

    Using Eq. 2 and assuming the dilute limit of defects, the formation energies of CPDVs and associated defect complexes are examined. Considering a scenario where CPDVs are introduced

into the graphene/*h*-BN system through irradiation, we outline the evolution possibilities of the system in Fig. 4a, where depending on the resultant defect complex of the lowest formation energy the 2D phase space of chemical environment and charge state is partitioned into different domains. Several key observations can be drawn from Fig. 4a. Firstly we note that in general the graphene/*h*-BN with CPDVs will always undergo the cross-planar migration process to transform into impurity doped graphene plus DV decorated *h*-BN (i.e., $C_X$ + $DV_{BN}$ with $X$ = B or N) or C doped *h*-BN plus DV decorated graphene (i.e., $X_C$ + $DV_C$ with $X$ = B or N). Secondly we see that the defect complexes in the left three domains (i.e., light purple and yellow domains) where the environment is largely B-rich derive from CPDVs with interlayer C-B links while those defect complexes in the right two domain where the environment is largely N-rich derive from CPDVs with interlayer C-N links, suggesting that the bonding nature at a CPDV can be controlled by varying the chemical potential of B or N. In addition, we see that the cross-planar migration route at a CPDV can be precisely regulated via the charge state. This is of particular significance as it enables unidirectional atom transport from *h*-BN to graphene to grant consistent doping and healing of graphene. Fig. 4a provides a predictive mapping of structural evolution for irradiated graphene/*h*-BN heterostructures, and suggests a novel approach, integrating irradiation, modulation of chemical potential and charge engineering to predictively functionalize graphene on top of *h*-BN, as schematically illustrated in Fig. 4b.

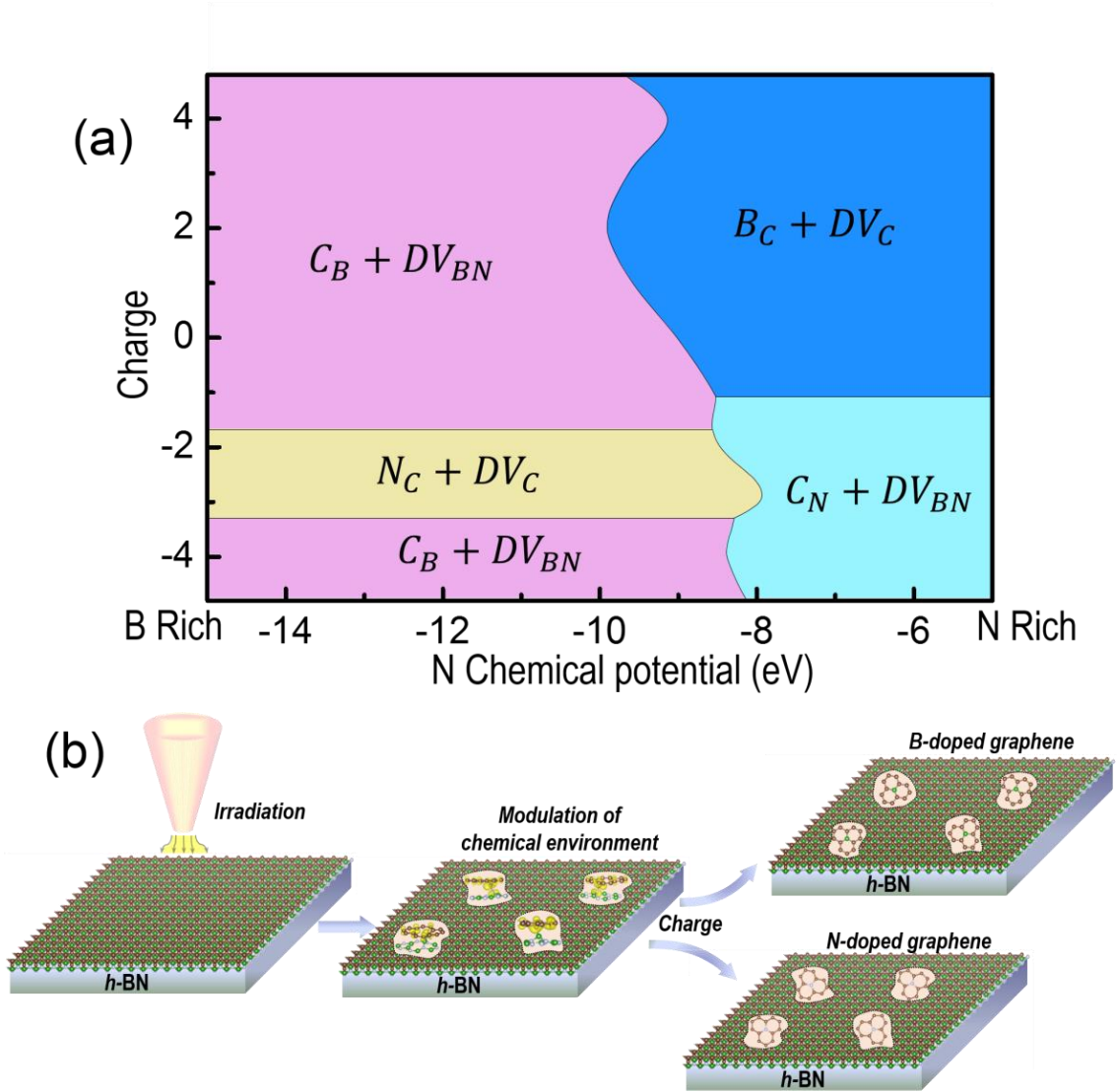

**Fig. 4** (Color online) (*a*) Predictive mapping of the evolution possibilities of irradiated stacked graphene/*h-BN* heterostructures as the chemical potential and charged state vary; (b) Schematic illustration of the possible experimental procedure to use electron or ion beam irritation to induce CPDVs in *h-BN* supported graphene, augmented by chemical potential and charge engineering to achieve controlled B or N doping of the graphene sheet.

To understand the pronounced effects of charge on structural evolution at CPDVs, we examine the formation energetics and electronic structures of charge doped CPDVs and their associated defect complexes. Fig. 5 shows the formation energy difference ($\Delta E_f^q$) and Fermi energy difference ($\Delta E_F^q$) of different structural evolution possibilities at two representative

CPDVs, i.e., $V_2^1(C_\psi N)$ in a B-rich environment[53] and $V_2^1(C_\varphi B)$ in a N-rich environment[54], as the charge state $q$ varies, with

$$\Delta E_f^q = E_f^q[TS] - E_f^q[V_2^i(C_\alpha X)], \qquad (3)$$

$$\Delta E_F^q = E_F^q[TS] - E_F^q[V_2^i(C_\alpha X)], \qquad (4)$$

where $E_f^q[TS]$ and $E_F^q[TS]$ respectively denote the formation energy and Fermi energy of a particular structural evolution configuration while $E_f^q[V_2^i(C_\alpha X)]$ and $E_F^q[V_2^i(C_\alpha X)]$ respectively denote the formation energy and Fermi energy of the corresponding CPDV $V_2^i(C_\alpha X)$ ($i$ = 1 or 2, $\alpha = \psi$ or $\varphi$, and $X$ = B or N). For the case of $V_2^1(C_\psi N)$, we can note from Fig. 5a that $B$ doping and simultaneous healing of graphene is energetically preferred when the system is doped with the charge state being $q \geq -1$ or $q < -3$ while the evolution towards $N_C$ + DV$_C$ is preferred otherwise. Meanwhile for the case of $V_2^1(C_\varphi B)$ (cf. Fig. 5d), $N$ doping and simultaneous healing of graphene is preferred when the system is negatively charged with $q < -1$ while the evolution towards $B_C$ + DV$_C$ is preferred otherwise.

Noting from Eq. 2 that the defect formation energy has an apparent dependence on the Fermi energy $E_F^q$, and combining Eqs 2-4, we have

$$\Delta E_f^q = -q\Delta E_F^q + (E_{tot}^q[TS] - E_{tot}^q[V_2^1(C_\alpha X)]), \qquad (5)$$

which further yields (denoting $\Delta E_{tot}^q = E_{tot}^q[TS] - E_{tot}^q[V_2^1(C_\alpha X)]$ for simplicity)

$$\frac{\partial \Delta E_f^q}{\partial q} = -\Delta E_F^q + \frac{\partial \Delta E_{tot}^q}{\partial q}. \qquad (6)$$

Eq. 6 demonstrates that the variation of $\Delta E_f^q$ with respect to $q$ explicitly depends on $\Delta E_F^q$ (albeit the presence of $\partial \Delta E_{tot}^q / \partial q$), which is also clearly evident from Fig. 5 (see Supplementary Information for detailed analysis).

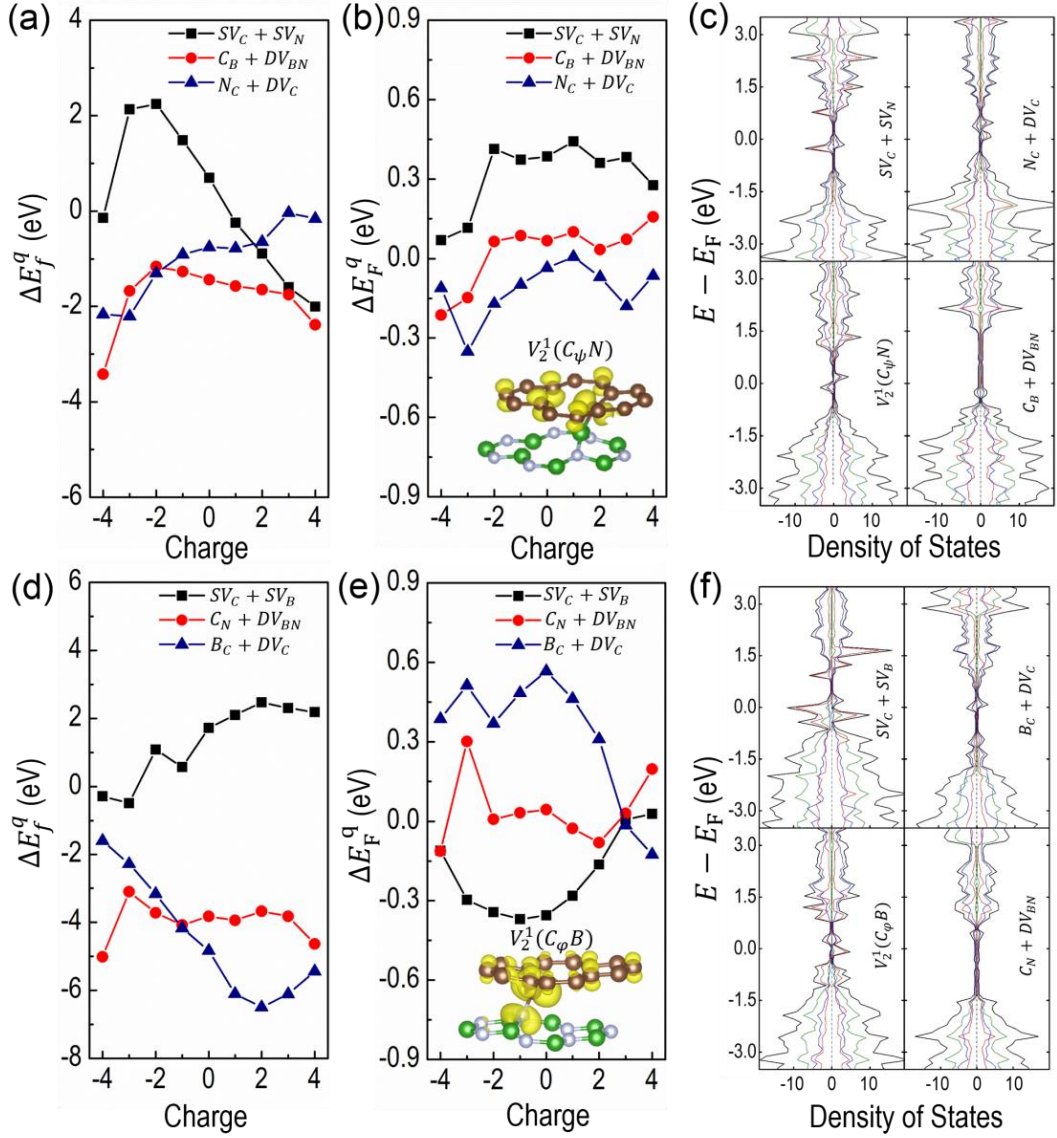

**Fig. 5** (Color online) (a) The formation energy difference $\Delta E_f^q$ (cf. Eq. 3) and (b) Fermi energy difference $\Delta E_F^q$ (cf. Eq. 4) of different structural evolution possibilities at $V_2^1(C_\psi N)$ as functions of the charge state $q$. (d) $\Delta E_f^q$ and (e) $\Delta E_F^q$ of different structural evolution possibilities at $V_2^1(C_\varphi B)$. Also shown are the projected density of states (PDOS) plotted for different defect complexes associated with (c) $V_2^1(C_\psi N)$ and (d) $V_2^1(C_\varphi B)$ respectively, with black, green, blue and red lines indicating the total density of states, PDOS from graphene, PDOS from $h$-BN and PDOS contributed by the corresponding impurity states. The insert figures in (b) and (e) illustrate the partial charge of $V_2^1(C_\psi N)$ and $V_2^1(C_\varphi B)$ only contributed by impurity states respectively.

Meanwhile the projected density of states (PDOS) for different defect complexes associated with $V_2^1(C_\psi N)$ and $V_2^1(C_\varphi B)$ are also plotted in Fig. 5. With the PDOS contributions from graphene, $h$-BN and defect impurity states individually indicated, it shows that the difference between $\Delta E_f^q$ of different defect complexes mainly come from the impurity states. For those possible defect complexes evolving from CPDVs, we can view them as collections of in-plane defect constituents (i.e., single or di-vacancies, substitutional impurities). Accordingly the Fermi level and band alignment near Fermi level[5, 55-57] is determined by the characteristics and interplay between the defect constituents involved[25, 26, 30, 33, 58-64]. For instance, the Fermi energy generally can be elevated in the presence of *n*-type impurities, e.g., $SV_N$[25, 26, 58], but decrease in the presence of *p*-type impurities, e.g., $SV_B$[26, 58, 60]. The dependence of the Fermi energy on the defect constituents thus differs with the introduction of charge, as the injected electrons or holes will occupy different energy states (See Supplemetary Information for details).

**Conclusions:**

In summary, we examined the cross-planar di-vacancies (CPDVs) at graphene/*h*-BN interface. Those extended defects consist of interlayer covalent bonds and act as effective channels for atom transport between adjacent graphene and *h*-BN sheets. When directed towards graphene, the CPDV-assisted cross-planar migration induces B or N doping and simultaneous removal of vacancies to heal the lattice structure of graphene. We showed that, by tuning the chemical environment and charge state, the chemical nature of the interlayer bond and associated cross-planar migration pathways at CPDVs can be manipulated to predictively grant consistent B or N doping and healing of graphene. Our findings suggest a viable experimental recipe, combining irradiation, chemical and charge engineering, to produce high-quality graphene with tunable electronic and electrochemical properties, using the *h*-BN substrate.

# Supplementary information

Bin Ouyang[1], J. Song[1, a)]

1. Department of Mining and Materials Engineering, McGill University, Montreal, QC, Canada

**Computational Methodology:**

Stacked bilayer graphene/$h$-BN heterostructures consisting of one layer of graphene and $h$-BN each are considered in the present study. In our simulations, an $8\times 8$ of unit cell is used. The simulation cell is shown to be large enough to eliminate interactions between defects and their periodic images. The lattices of graphene and $h$-BN are set to be the same, being $a_0$=2.49Å which is found to give the lowest total energy. The dimension of the vacuum space perpendicular to the bilayer heterostructure is set as 15Å to avoid image interaction[1]. There are three possible stable stacking configurations for bilayer graphene/$h$-BN stacking with slight stacking energy difference. In this paper, the stacking with nitrogen atom on top of the hexagonal center of carbon (AB-GBN) is selected as it represents the most stable stacking configuration (Fig. S1a). Lattice defects are then introduced into the bilayer structures to construct different defect complexes.

Spin polarized DFT calculations were performed using the Vienna ab-initio Simulation Package (VASP)[2] with projector augmented-wave (PAW)[3-7] pseudopotentials. A cutoff energy of the plane wave basis set of 500 eV is used in all calculations. Further increase in the cutoff energy up to 800eV will only introduce a tiny energy difference < 0.02eV. The climbed image Nudged Elastic Band (ci-NEB) method is employed to calculate minimum reaction paths (MEPs)

---

a) Author to whom correspondence should be addressed. E-Mail: jun.song2@mcgill.ca

of migration processes[8-11]. Both the structure optimization and ci-NEB calculations are regarded converged when the force is < 0.01 eV/Å.

**Atomic Configurations and Simulated STM images for Other CPDVs:**

There are in total eight possible CPDVs in AB-GBN. Besides the four representative CPDVs shown in Fig.2, the other four configurations along with the simulated STM images are illustrated in Fig. S1 below, which (together with Fig. 2 in main text) gives a complete picture of the ground states of CPDVs in AB-GBN and how them and their features under STM imaging.

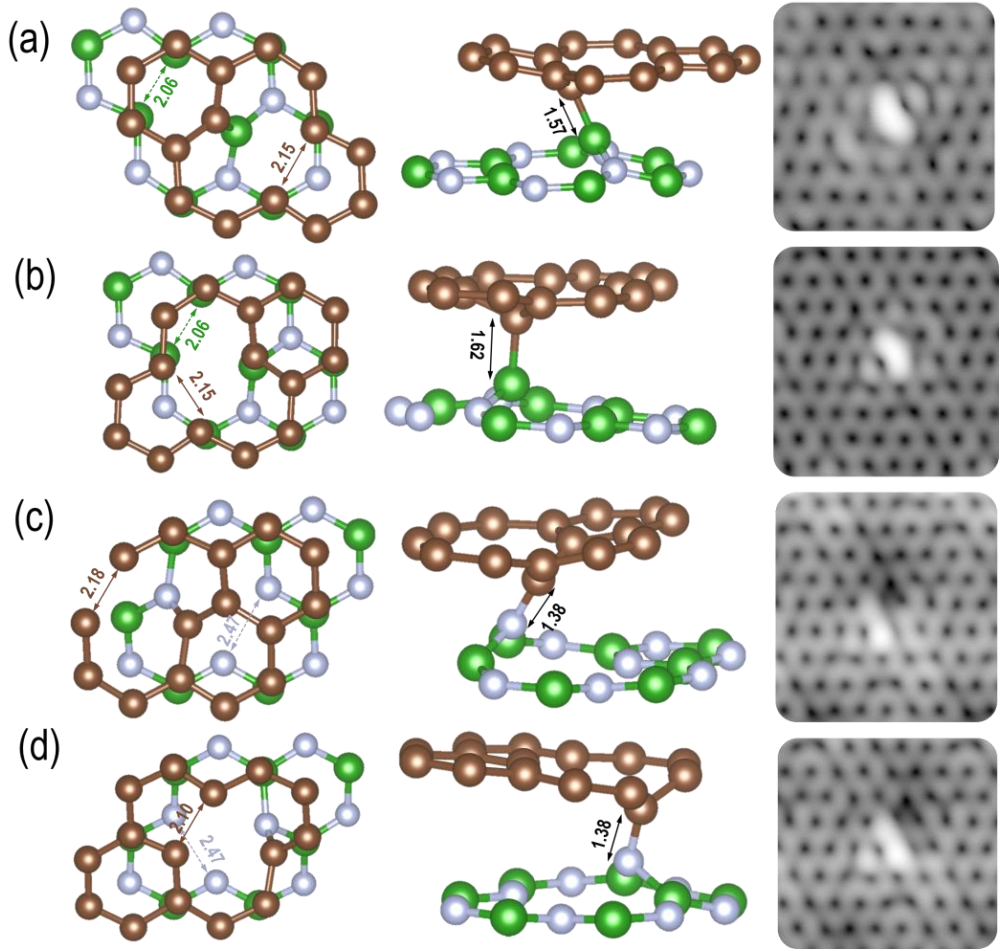

**Fig. S1** (Color online). The top and side views of the ground-state configurations of CPDVs: (a) $V_2^2(C_\varphi N)$, (b) $V_2^1(C_\varphi N)$, (c) $V_2^2(C_\Psi N)$ and (d) $V_2^1(C_\Psi N)$. The subfigures in the rightmost column are the simulated STM images of those CPDVs. In the STM simulation, the partial charge range from $E_F - 1.5$eV to $E_F$ (with $E_F$ being the corresponding Fermi energy) is considered to capture the defective charge state. C, B and N atoms are colored dark brown, green and silver respectively.

**Energy Release and Basal Shifting from CPDVs**

The energy release associated with the formation of a CPDV comes from the bond forming and local lattice distortion. One apparent aspect in the CPDV-induced lattice distortion is the basal shifting, along either zigzag (ZZ) or armchair (AC) directions, or both, as illustrated in Fig. 2 as well as Fig. S2 below. The basal shifting modifies the local stacking between graphene and *h*-BN sheets and thus requires energy. Fig. S1 shows the generalized stacking fault energy profiles, defined according to Telling et al.[12], for basal shifting along ZZ and AC directions in perfect AB-GBN, clearing illustrating the dependence of energy cost on the shifting direction. In particular, for basal shifting of small magnitude (i.e., < 0.22 $a_0$ or 0.55Å, which is generally the case for a CPDV-induced shifting), we can note that it is more difficult for the shifting to occur along AC than ZZ direction. Thus the data presented in Fig. S1 can help explain the difference between formation energies of different CPDVs.

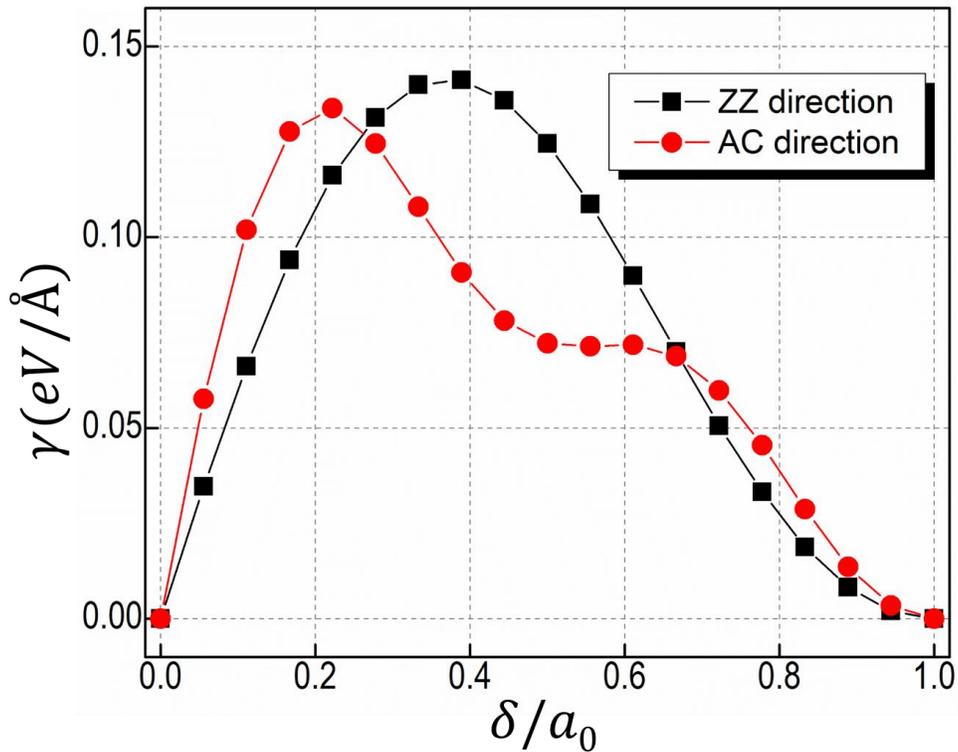

**Fig. S2** (Color online) Generalized stacking fault energy profiles for basal shifting along ZZ and AC directions.

**Influence of Charge on Fermi Energies and Formation Energies**

The Fermi energies and formation energies of CPDVs are calculated to further help understand the structural evolution at CPDVs as the charge state varies. Fig. S3 shows the results for two representative CPDVs, i.e. $V_2^1(C_\psi N)$ in a B-rich environment[20] and $V_2^1(C_\varphi B)$ in a N-rich environment[21]. This supplements Fig. 5 in the main text. We can see from Fig. S3 that as the charge state varies from the negative end to positive end, for all defect complexes the Fermi energy $E_F^q$ will increase while meanwhile the formation energy $E_f^q$ will first decrease and then begin to increase as the charge state is approaching the neutral state.

As seen in Eq. 2, there is an explicit dependence of $E_f^q$ on $E_F^q$. Using the corresponding CPDV as the baseline state, we examined $\Delta E_f^q$ (effective the energy release associated with the relevant structural evolution at CPDV), the variation of which with respect to $q$, i.e., $\partial \Delta E_f^q / \partial q$, directly comes from $-\Delta E_F^q$ plus the contribution from $\partial \Delta E_{tot}^q / \partial q$. We can also see from Fig. 5 that in certain cases $-\Delta E_F^q$ plays a dominate role in determining $\partial \Delta E_f^q / \partial q$. For instance, for the case of $C_B + DV_{BN}$, $\Delta E_f^q$ exhibits a positive slope for $-4 < q < -2$ but a negative slope for $q > -2$, and in accordance $-\Delta E_F^q$ is positive for $-4 < q < -2$ but a negative slope for $q > -2$. For the case of $SV_C + SV_N$, $\partial \Delta E_f^q / \partial q$ shows a nearly constant slope ~ -0.66 eV $-2 < q < 4$ while $-\Delta E_F^q$ stays close to -0.45 eV.

It is apparent from the DOS plots that $\Delta E_F^q$ derives from the impurity states. In particular, for all those defect complexes evolving from CPDVs, they can be viewed as collections of in-planar defects, which can generally be grouped into either $p$-type defects ($C_B$[22], $N_C$[23-25], $DV_{BN}$[17, 18], $SV_B$[17, 18, 26], $SV_C$[27, 28] $DV_C$[14]) or $n$-type defects ($C_N$[29], $B_C$[23-25], $SV_N$[17, 18, 25]). Consequently for a defect complex, the Fermi level and band alignment near Fermi level[30-33] are determined by the collabration of the in-plane defects involved. The influence of charge on $\Delta E_F^q$ can be understood by accounting for the details of the injected electrons or holes occupying different energy states in different defect complexes.

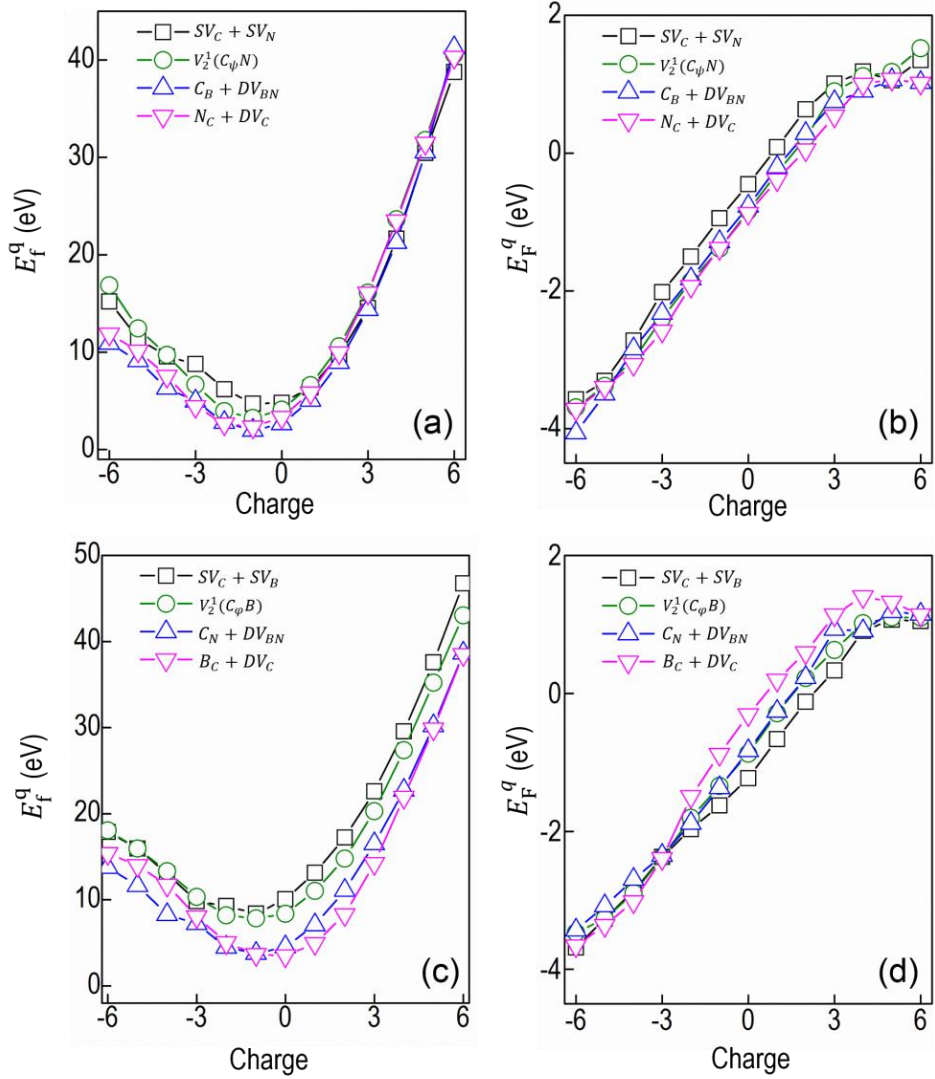

**Fig. S3** (Color online) The evolution of Fermi energy $E_F^q$ and defect formation energy $E_f^q$ as the charge state varies for (a) Evolution of defect formation energy for $V_2^1(C_\psi N)$ and correspondent transitional states; (b) Evolution of Fermi energy for $V_2^1(C_\psi N)$ and correspondent transitional states; (c) Evolution of defect formation energy for $V_2^1(C_\varphi B)$ and correspondent transitional states. (d) Evolution of Fermi energy for $V_2^1(C_\varphi B)$ and correspondent transitional states;